\documentclass[
]{ceurart}

\usepackage{listings}
\usepackage{url}

\def\showcomments{}
\ifdef{\showcomments}
{

}

\newcommand{\eqdef}{\overset{\tiny{def}}{=}}
\begin{document}

\copyrightyear{2022}
\copyrightclause{Copyright for this paper by its authors.
  Use permitted under Creative Commons License Attribution 4.0
  International (CC BY 4.0).}

\conference{}

\title{Unsupervised Search Algorithm Configuration using Query Performance Prediction}


\author{Haggai Roitman}[%
email=hroitman@ebay.com,
]
\cormark[1]
\address{eBay Research,
  Netanya, Israel}



\cortext[1]{Work was done while the author was still in IBM Research}

\begin{abstract}
  Search engine configuration can be quite difficult for inexpert developers. Instead, an auto-configuration approach can be used to speed up development time. Yet, such an automatic process usually requires relevance labels to train a supervised model. In this work, we suggest a simple solution based on query performance prediction that requires no relevance labels but only a sample of queries in a given domain. Using two example usecases we demonstrate the merits of our solution. 
\end{abstract}


\maketitle

\section{Introduction}\label{sec:intro}

Search engine solutions such as the (Apache) Lucene\footnote{\url{https://lucene.apache.org/}} library and Elasticsearch\footnote{\url{https://www.elastic.co/elasticsearch/}} have become a ubiquitous utility for developers of various discovery and data mining applications. Yet, configuring such solutions for a specific application in mind can be quite challenging and time-consuming. Commonly, inexpert developers may find it hard to determine which configuration would best fit their needs. Therefore, many such developers usually prefer to use the ``out-of-the-box'' default configuration (e.g., BM25 similarity in Elasticsearch). 

Nowadays, contemporary search engine solutions offer a variety of alternative configurations that can be utilized to improve search effectiveness. As a motivating usecase (which is also studied later on in our evaluation), the Divergence From Randomness~\cite{DFR} (DFR) similarity, one of the many similarities implemented within the Lucence search library, has more than 100 different configurations. Therefore, even for a single search algorithm option, it is highly difficult for developers to determine in-advance which DFR configuration is best for their application.

Configuring the search algorithm of a search engine so as to optimize its search effectiveness for a given application in mind usually requires some sort of domain expertise. In addition, choosing the ``best'' configuration usually requires enough relevance labels to train a supervised auto-configuration model~\cite{auto-tune}, which are not always in the expense of developers. 

In this work, we propose an \textbf{unsupervised} automatic search engine search algorithm configuration solution based on query performance prediction~\cite{qpp-tutorial} (QPP). We formally define the search algorithm configuration task as a utility maximization task, where the goal is to maximize the relevance of documents retrieved from the search engine for a given set of ``representative'' user-queries in the domain of interest. To estimate the expected relevance that may be obtained by a given candidate (search algorithm) configuration, we propose a simple, yet highly effective, extension to the probabilistic QPP framework~\cite{pQPP}.

Our solution only requires a sample of queries in the domain (e.g., query log) and a set of candidate search algorithm configurations to choose from; while no other input such as relevance labels is required. 
As a proof of concept, we evaluate our proposed solution on two common search algorithm configuration usecases in Lucene, namely: \emph{Similarity model selection} and \emph{Similarity parameter auto-tuning}.  


\section{Automatic Configuration Solution}
We now describe our automatic-configuration solution. To this end, we assume a corpus of documents $C$ which is searchable through a given search engine interface (e.g., Lucene library or Elasticsearch service). To derive an ``optimal'' search algorithm configuration over $C$, all we require is a sample of queries $Q$ (hereinafter termed the ``query benchmark'') and a set of (search algorithm) configurations $\mathcal{S}=\{S_1,\ldots,S_m\}$ to be evaluated. Our goal is, therefore, to find a configuration $S\in{\mathcal{S}}$ that would ``maximize'' the search engine's effectiveness. For each query $q\in{Q}$, the retrieval quality is expected to be measured relatively to the relevance of documents in $C$ retrieved (and ranked) according to $S$. Yet, in this work, we assume that relevance labels are unavailable apriori; instead, we aim to estimate such relevance using a QPP approach.  

Our solution builds on top of the \emph{probabilistic} QPP framework~\cite{pQPP}, introducing a simple, yet highly effective, extension in which instead of considering a single query $q\in{Q}$ effectiveness to be predicted, we wish to derive a prediction for the whole benchmark $Q$. 
Formally, let $r$ denote the relevance event. For a given configuration $S$, our goal is to estimate\footnote{We denote by $p(\cdot)$ a likelihood term and by $\hat{p}(\cdot)$ its corresponding estimate, which is \textbf{not always normalized}.} $p(r|Q,S,C)$ -- \emph{the  relevance likelihood given that we retrieve documents from $C$ for each query in $Q$ using $S$}. 

We now derive $p(r|Q,S,C)$ by first conditioning over the various queries $q$ in the benchmark:
\vspace{-0.1in}
\begin{equation}\label{eq:main estimate}
  p(r|Q,S,C)\eqdef \sum\limits_{q\in{Q}}p(r|q,S,C)p(q|Q),
\end{equation}
where $p(r|q,S,C)$ denotes the relevance likelihood for a specific query $q\in{Q}$ and $p(q|Q)$ denotes the likelihood of observing query $q$. We next estimate these two likelihood terms. 

\subsection{Estimating $p(r|q,S,C)$}\label{first term}
Estimating the performance of a specific configuration $S$ can be done by evaluating that configuration over $C$ using query $q$ and obtaining the response (result-list) $D_{S}^{q}\subseteq{C}$. Hence, in this work, we simply estimate $\hat{p}(r|q,S,C)\eqdef p(r|q,D_{S}^{q},C)$. 
To estimate $p(r|q,D_{S}^{q},C)$, we first note that: $p(r|q,D_{S}^{q},C)\propto p(r|q,D_{S}^{q})p(r|D_{S}^{q},C)$.
Here we further note that, while the left term $p(r|q,D_{S}^{q})$ aims to capture the \emph{query-dependent} quality effects of documents in $D_{S}^{q}$, the right term $p(r|D_{S}^{q},C)$ aims to capture the \emph{corpus-dependent} quality effects. 
Conditioning on the retrieved documents in $D_{S}^{q}$, we now extend the two terms as follows:

$p(r|q,D_{S}^{q})=\sum\limits_{d\in{D_{S}^{q}}}p(r|q,d)p(d|q,D_{S}^{q})$ and $p(r|D_{S}^{q},C)=\sum\limits_{d\in{D_{S}^{q}}}p(r|d,C)p(d|D_{S}^{q})$. 

Finally, we derive the remaining likelihood terms as follows. 
We first assume that the query-independent term $\hat{p}(d|D_{S}^{q})$ is uniform over $D_{S}^{q}$. Next, we estimate the query-dependent term $\hat{p}(d|q,D_{S}^{q})\eqdef\frac{S(d|q)}{\sum_{d'\in{D_{q}^{l}}}S(d'|q)}$, where $S(d|q)$ denotes the retrieval score of document $d$ assigned by applying configuration $S$ to evaluate query $q$. Following~\cite{WPMQPP}, we estimate $p(r|q,d)$ according to BM25 score. Finally, further following~\cite{WPMQPP}, we estimate $p(r|d,C)$ according to document $d$'s ``focus'' in $C$, as follows: $\hat{p}(r|d,C)\eqdef \exp\left(-\sum\limits_{w\in{d}} p(w|d)\log \frac{p(w|d)}{nidf(w)}\right)$, 
where $p(w|d)$ denotes term $w$'s likelihood in $d$'s language model and $nidf(w)=\frac{idf(w)}{\sum\limits_{w'\in{d}}idf(w')}$.

\subsection{Estimating $p(q|Q)$}
The simplest way to estimate $p(q|Q)$ is to assume that queries in $Q$ are independent of each other, and hence, we could completely ignore this term. Yet, $p(q|Q)$ can be interpreted with other important meanings. For example, $p(q|Q)$ may represent the popularity of $q$ in the domain being searched. An alternative (and novel) estimation, which we shall shortly derive, is based on the relative ``difficulty'' of $q$ compared to other queries in the benchmark. In such a case, we assume that a more difficult query is more important. This in turn allows to quantify the potential of a certain configuration to provide relevant information in a variety of query ``difficulty levels'' that may exist in a certain domain. Hence, we would like to bias our configuration selection solution toward configurations that better handle more ``difficult'' queries.

To quantify a given query's difficulty, we now utilize a \emph{voting} approach, where we first estimate the relevance log-likelihood of each query $q\in{Q}$, assuming all configurations in $\mathcal{S}$ are used to serve that query, i.e.: 
\vspace{-0.2in}
\begin{equation}
p(r|q,\mathcal{S},C)\eqdef \log\prod\limits_{S\in\mathcal{S}}p(r|q,S,C),
\end{equation}

where $p(r|q,S,C)$ can be realized by any QPP method (e.g., similar to the one we have just derived in Section~\ref{first term}). For more practical reasons, wishing to enhance the ``relevance signal'', we suggest to utilize a much simple approach. In our case, we use the NQC method~\cite{NQC}.

We next sort queries in $Q$ according to  $p(r|q,\mathcal{S},C)$. Let $q_{i}$ denote the query $q\in{Q}$ ranked at the $i$-th position ($1\leq{i}\leq{|Q|}$). We then estimate $p(q_{i}|Q);1\leq{i}\leq{|Q|}$, as follows:
\vspace{-0.1in}
\begin{equation}\label{eq:query diff}
\hat{p}(q_{i}|Q)\eqdef 1 - \frac{\sum\limits_{j<i}p(r|q_{j},\mathcal{S},C)}{\sum\limits_{q\in{Q}}p(r|q,\mathcal{S},C)}
\end{equation} 

\begin{figure}
    \centering
    \includegraphics[width=0.8\textwidth]{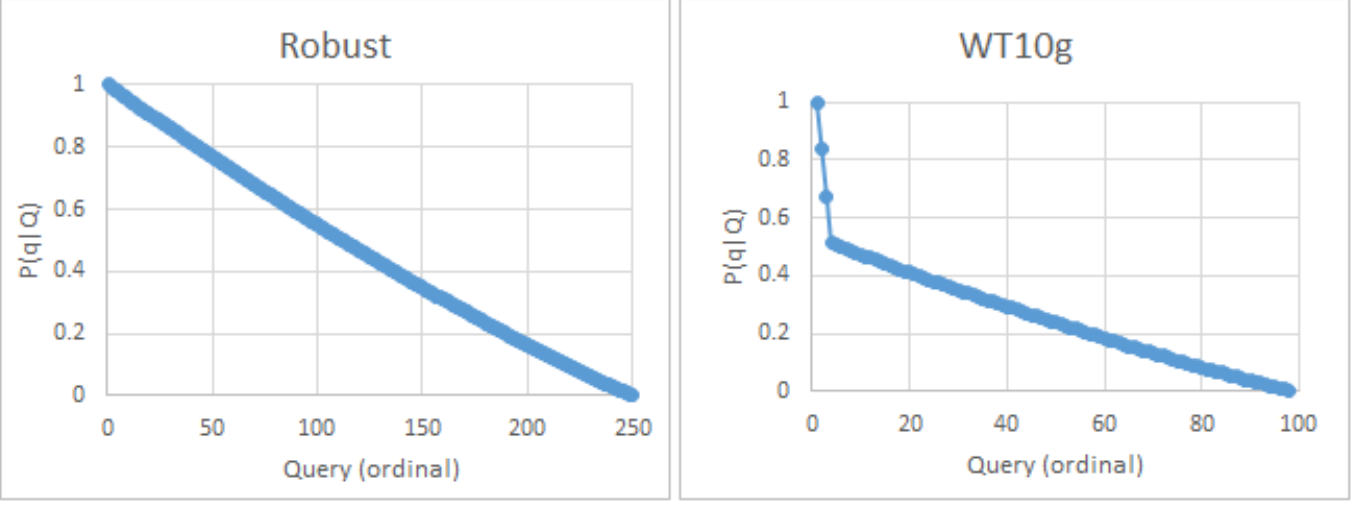}
    \caption{Query difficulty estimates for the Robust and WT10g benchmarks. A higher value, denotes a more difficult query.}
    \label{fig:query weights}
\end{figure}

As illustrative examples, Figure~\ref{fig:query weights} depicts the $p(q|Q)$ estimated values according to Eq.~\ref{eq:query diff} for the Robust and WT10g query benchmarks. As one would expect, the Robust benchmark has queries of diverse difficulties (based on its intended purpose); and indeed, the $p(q|Q)$ values follow a linear trend. On the other hand, the WT10g benchmark has much fewer difficult queries, and as we can observe, only those receive high values. This serves as an empirical evidence to our design choice of the $p(q|Q)$ estimate, where we wish to prefer configurations that can handle more difficult queries of a given benchmark.

\section{Evaluation}

We evaluate our proposed unsupervised auto-configuration solution using several ad-hoc document retrieval benchmarks. To this end, we use the Robust, WT10g, AP, WSJ and SJMN datasets.  As a proof of concept, we evaluate our proposed solution on two common search algorithm configuration usecases in Lucene, namely: \emph{Similarity model selection} and \emph{parameter auto-tuning of a given Similarity}. We use Lucene's standard English text analysis pipeline. 

In this work, we use MAP as our target relevance metric. To this end, for each configuration $S\in\mathcal{S}$ and for each query $q\in{Q}$, we retrieve the top-$100$ documents in $C$ with the highest retrieval score according to $S$.
Using a \textbf{post-analysis} evaluation approach, given a set of configurations $\mathcal{S}$ to explore and benchmark $(Q,C)$, we choose the best configuration in $S^{*}\in\mathcal{S}$ according to Eq.~\ref{eq:main estimate}. We then measure MAP obtained by each configuration, and use the highest MAP as the performance upper-bound. We next define MAP$_{lift}(i,j)\eqdef\frac{MAP(S_i)}{MAP(S_j)}$ as the relative lift in MAP obtained by using configuration $S_i$ instead of $S_j$.
As our first quality metric, we measure the lift of our selected configuration $S^{*}$ compared to the optimal selection $S^{opt}\eqdef\arg\max_{S\in\mathcal{S}}MAP(S)$. We further measure the lift compared to the hypothetical random configuration selection, with $MAP(random)\eqdef\frac{1}{|\mathcal{S}|}\sum\limits_{S\in\mathcal{S}}MAP(S)$. Finally, we evaluate how well our solution is able to rank the different configurations by measuring the correspondence (measured by kendall-tau rank-correlation) between the order obtained according to Eq.~\ref{eq:main estimate} and the one obtained using the actual post-analysis MAP values. 

\subsection{Usecase 1: Similarity model selection}

\begin{table}[t!]
\caption{Evaluation results for the Similarity model selection usecase.}
\begin{tabular}{l|c|c|c|c|c|}
\cline{2-6}
\multicolumn{1}{c|}{}               & \textbf{Robust} & \textbf{WT10g} & \textbf{AP} & \textbf{WSJ} & \textbf{SJMN} \\ \hline
\multicolumn{1}{|l|}{MAP$_{lift}(S^{*},S^{opt})$}  & .994            & .999           & .992        & 1.00         & .981          \\ \hline
\multicolumn{1}{|l|}{MAP$_{lift}(S^{*},S^{random})$} & 1.067           & 1.077          & 1.042       & 1.092        & 1.052         \\ \hline
\multicolumn{1}{|l|}{KT-correlation}      & 0.667           & 0.733          & 0.867       & 0.667        & 0.733         \\ \hline
\end{tabular}\label{tab:similarities}
\end{table}

In this usecase, we wish to demonstrate the ability of our auto-configuration solution to choose amongst several \texttt{Similarity} models supported by Lucene. To this end, we use four popular \texttt{Similarity} model implementations, namely: BM25, LM (with Dirichlet smoothing), DFR and Information-Based (IB). For each \texttt{Similarity} model we consider the \textbf{default} parameters that are already configured in Lucene. We further extend the configuration set with two \texttt{Rescorer} (reranking) implementations using pseudo relevance models, namely RM1 and RM3. 
To this end, we first retrieve 100 documents using the LM similarity model, and then rerank documents based on the induced relevance model (RM). 

We summarize the results of our evaluation in Table~\ref{tab:similarities}. First, as we can observe, for all datasets, the configuration that we select is almost similar to the optimal configuration. Since there are only six possible configurations in this usecase, for any lift value that is below $1$ that would probably mean that we usually select the second best configuration (except for WSJ where we selected the best one). Further comparing our performance to the random selection, we can observe that, our lift is always above $1$, meaning we find a configuration that is always better than the random choice. Our selection provides at least $+4\%$ improvement in MAP (with an average of $+6\%$). Finally, the KT-correlation values we obtain are high, meaning that our selection order mostly agrees with the optimal selection order.

\subsection{Usecase 2: Similarity parameter auto-tuning}

\begin{figure}[th!]
    \centering
    \includegraphics[width=0.8\textwidth]{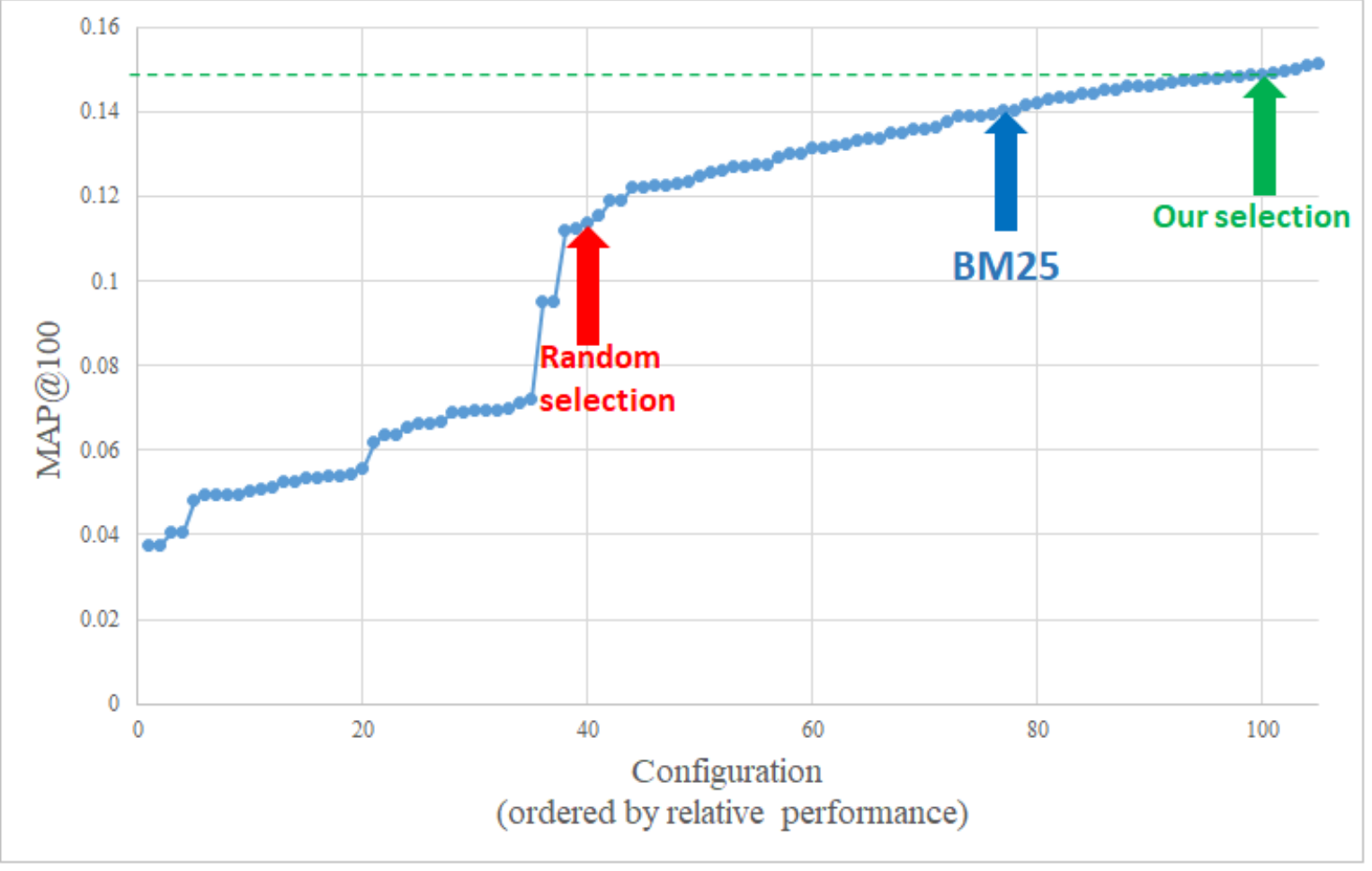}
    \caption{Evaluation results of the Similarity parameter auto-tuning usecase.}
    \label{fig:parameter}
\end{figure}

In this usecase, we wish to demonstrate the ability of our auto-configuration solution to choose amongst several parameter configurations of a given \texttt{Similarity} model. This is a common usecase for Lucene developers who usually cannot know in advance which parameter configuration would best fit their needs; hence, such developers usually prefer to use the default \texttt{Similarity} configuration of Lucene (in this case BM25). 

To make an interesting usecase, we wish to answer the following two questions: ``Would it be better to use the DFR Similarity instead of BM25?'' and ``Given that we choose the DFR Similarity, how good would be our selection?''. To answer both questions, we now use the Robust dataset and evaluate the BM25 Similarity together with all possible DFR Similarity configurations (overall 105 possible configurations\footnote{7 \texttt{BaseModel} options, 3 \texttt{AfterEffect} options and 5 \texttt{Normalization} options.}).  

Figure~\ref{fig:parameter} depicts the MAP values obtained by the BM25 configuration and all DFR configurations. We further order configurations relative the MAP values they obtained. As we can first observe, choosing the DFR Similarity over BM25 for the Robust dataset can be indeed better, but depends on what configuration is being used (with only about 30 configurations that are more promising). Overall, our solution has chosen a DFR configuration that yeilds a significant lift in MAP over that of BM25 (about $+7\%$ improvement). Moreover, as we can observe, a random choice\footnote{We note that the default DFR similarity of Lucene has performance similar to BM25 for this dataset.} in this case, results in a sub-optimal quality (with about $+30\%$ lift obtained by our solution), which again demonstrates the effectiveness of our more ``educated'' selection. 

\section{Conclusions}
We proposed a simple, yet highly effective, unsupervised search algorithm configuration selection solution. Using our solution, a developer can be assisted in auto-tuning a configuration that better fits her needs by only providing a sample of queries in the domain of interest. While we only presented two example usecases, we have also successfully evaluated the solution on more complex usecases such as the possibility to combine several search algorithms.  


\bibliography{qpp}




\end{document}